\documentclass[twocolumn,aps,prc,showpacs,floatfix]{revtex4}
\usepackage{graphicx}
\usepackage{amsmath}
\usepackage{dcolumn}
\usepackage{bm}

\begin{document}

\title{Probing nuclear bubble configuration by the $\pi^{-}/\pi^{+}$ ratio in heavy-ion collisions}

\author{Gao-Chan Yong}

\affiliation{%
{Institute of Modern Physics, Chinese Academy of Sciences, Lanzhou
730000, China}
}%

\begin{abstract}
It is theoretically and experimentally argued that there may exist
bubble or toroid-shaped configurations in some nucleus systems. Based on the isospin-dependent
transport model of nucleus-nucleus collisions, here we propose a method to probe
the bubble configuration in nucleus. That is, one could use the
value of $\pi^{-}/\pi^{+}$ ratio especially its kinetic energy
distribution in head-on collision at intermediate energies to probe whether there is bubble configuration or not in projectile and target nuclei.
Due to different maximum compressions and the effect of symmetry energy, the value of $\pi^{-}/\pi^{+}$ ratio in the collision of bubble nuclei is evidently larger than that in the collision of normal nuclei.

\end{abstract}

 \maketitle

\section{Introduction}

It was argued that a nuclear system
may have bubble configuration
\cite{Decharg99,Nazarewicz02,Decharg03}, which is in conflict with the general knowledge that an atomic nucleus is compact.
The question of the possibility of the existences of exotic bubble
or toroidal configuration of atomic
nuclei has long been discussed for more than 60 years
\cite{Wilson46,Siemens67,wong721,wong722,wong723,wong73,wongrmp,Campi73,Yu00}.
The pioneer study of the spherical bubble nuclei was made by Wilson in 1946 \cite{Wilson46}.
Twenty years later, based on a liquid drop model, Siemens and Bethe studied spherical bubble nuclei \cite{Siemens67}. Wong studied known stable nuclei and found spherical bubbles on the basis of a liquid drop model plus a shell correction energy \cite{wong73}.
And Moretto \emph{et al.} argued that nuclear bubble configuration could be stabilized by the inner vapor pressure \cite{more97}. Based on the hydrodynamic equations, Borunda and L\'{o}pez argued that the hollow configuration is caused by the liquid-like behavior of compressed nuclear matter \cite{borun94}. The transport model simulations of nuclear collisions at lower incident beam energy also point out the possible bubble configuration
in nuclear matter \cite{bauer92,liba92,liba93,xu93,zhang14,yongbf15}.
And J. Decharg\'{e} \emph{et al.} recently gave stable bubble solutions
of some specific super-heavy nuclei using the approach of the self-consistent microscopic Hartree--Fock--Bogoliubov calculations with the effective Gogny interaction \cite{Decharg99,Decharg03}.
These results are qualitatively consistent with the studies
based on the liquid drop model using Strutinsky shell correction
method \cite{wongrmp,swia66} plus the phenomenological shell model potentials \cite{Dietrich76,Dietrich98}. At the time being, the possible proton bubble configuration of some light nuclei has also been studied \cite{yaojm2013}.

The probe of the bubble configuration formed
in heavy-ion collisions has recently been argued in Refs.~\cite{bauer92,liba92,liba93,xu93,zhang14,yongbf15}, while the hollow configuration of atomic nucleus is in fact hard to probe \cite{najman2015}. This is the reason why the existence of bubble nuclei is still
in debate. Besides the possible electronic
hyperfine configuration measurement of the nuclear density profile \cite{Decharg03,suda2009,ant2011}, it is necessary to study whether
hollow atomic nuclei exist or not by other completely different methods.
Very recently, \emph{Najman et al.} analyzed the Data from an experiment on the $^{197}$Au+$^{197}$Au reaction at 23 MeV/nucleon and showed that
the exotic nuclear configurations such as toroid-shaped objects may exist \cite{najman2015}. This experimental investigation stimulates further studies on the exotic nuclear configurations theoretically.
Based on the isospin-dependent Boltzmann
nuclear transport model, here we show that the observable
$\pi^{-}/\pi^{+}$ ratio in heavy-ion collisions at intermediate
energies is very sensitive to the initial nuclear bubble configuration. In the
following, for the convenience of studying, we use the bubble nucleus $^{900}_{274}$X as projectile and target to collide at a beam energy of 0.4
GeV/nucleon to show how to probe the nuclear bubble configuration in nucleus.
This example is fictitious, but it could demonstrate the point.

\section{The isospin-dependent transport model}

In the used isospin-dependent Boltzmann-Uehling-Uhlenbeck (BUU)
transport model, nucleon coordinates in initial colliding nuclei
are given by \cite{bertsch}
\begin{equation*} \label{xyz}
r = R(x_{1})^{1/3}; cos\theta = 1-2x_{2}; \phi = 2\pi x_{3},
\end{equation*}
\begin{equation}\label{1}
    x = rsin\theta cos\phi;
    y = rsin\theta sin\phi;
    z = rcos\theta.
\end{equation}
Where $R$ is the radius of compact nucleus, $x_{1}, x_{2}, x_{3}$ are
three independent random numbers (limited to be between 0 and 1). If there is a spherical bubble with radius $R_{bubble}$ in a nucleus, the radius $R_{b}$ of the nucleus with
bubble configuration is given by conservation of volume
\begin{equation}
R_{b}^{3} = R^3 + R_{bubble}^3.
\end{equation}
In the present study we let inner bubble radii $R_{bubble}$ =
$R/2, R/3$ \cite{Yu00}. Due to the short-range correlations of nucleons in nucleus, above Fermi momentum, momentum distribution of nucleon in nucleus exhibits a 1/$k^4$ high-momentum tail shape \cite{sci14}, the isospin-dependent nucleon momentum distribution in this study is thus given by
\begin{eqnarray}
n_{proton (neutron)}(p)=\left\{%
\begin{array}{ll}
    C_{1,proton (neutron)}, & \hbox{$p \leq p_{f}$;} \\
    \frac{A}{2Z (2N)}\cdot \frac{C_{2}}{p^{4}}, & \hbox{$p_{f} < p \leq \lambda p_{f}$;} \\
    0, & \hbox{$p > \lambda p_{f}$.} \\
\end{array}%
\right.
\end{eqnarray}
Where $n(p)$ is nucleon momentum distribution in initial colliding
nucleus, $p_{f}$ is the nuclear Fermi momentum.
$\lambda=p_{max}/p_{f}$, is the factor of maximum momentum of
nucleon relative to the nuclear Fermi momentum. We in this study
let $\lambda=2$ \cite{yong2015}. $A, Z, N$ are the nuclear mass
number, proton number and neutron number, respectively. The
$C_{1}$ and $C_{2}$ parameters are determined by the normalization
condition $4\pi\int_0^{\infty} n(p) p^2dp =1$ and roughly 20 percent of
nucleons with momenta above the nuclear Fermi momentum \cite{sci08}, i.e., $
4\pi\int_{p_f}^{\lambda p_{f}} n(p) p^2dp \times 100\% = 20\%. $
The 20 percent of nucleons with high momenta are randomly
distributed in the nucleus.

We use the Skyrme-type parametrization for the mean field, which
reads \cite{bertsch}
\begin{equation}
U(\rho)=A(\rho/\rho_{0})+B(\rho/\rho_{0})^{\sigma}.
\end{equation}
Where $\sigma = 1.3$, A = -232 MeV accounts for attractive potential and B =
179 MeV accounts for repulsive. With these choices, the
ground-state compressibility coefficient of nuclear matter is K = 230
MeV \cite{todd2005}. We let the kinetic symmetry-energy (the symmetry energy can be divided into the kinetic symmetry energy and the potential symmetry energy) be -6.71 MeV \cite{cli15}, and the
symmetry-potential becomes \cite{henli14}
\begin{equation}\label{usym}
U^{n/p}_{\rm sym}(\rho,\delta)=38.31(\rho/\rho_0)^{\gamma} \times
[\pm 2\delta+(\gamma-1)\delta^2],
\end{equation}
where $\delta=(\rho_n-\rho_p)/\rho$ is the isospin asymmetry of
nuclear medium. In the above, we let the value of the symmetry-energy
at saturation density be 31.6 MeV \cite{esym14,cxu10}. In the present study, we let $\gamma$ = 0.3 \cite{yong20152} for the soft symmetry energy choice while
$\gamma$ = 1.5 for stiffer symmetry energy choice as comparison.

The in-medium baryon-baryon ($BB$) elastic cross sections are
factorized as the product of a medium correction factor and the
free baryon-baryon scattering cross sections \cite{liq2006}, i.e.,
\begin{equation}
\sigma_{medium}^{BB,elastic}=(\frac{1}{3}+\frac{2}{3}e^{-u/0.54568})\times
(1\pm0.85\delta)\times\sigma_{free}^{BB,elastic}. \label{elastic}
\end{equation}
$u=\rho/\rho_{0}$ is the relative density. $1\pm0.85\delta$ is the isospin dependent factor.
Because pion's production and absorption are in fact mainly
determined by the inelastic baryon-baryon scattering cross
section, the in-medium correction of the inelastic baryon-baryon
scattering cross section has also to be taken into account. In
this model, for the inelastic baryon-baryon scattering cross
section, we use the form \cite{ko15}
\begin{equation}
\sigma_{medium}^{BB,inelastic}=(e^{-1.3u})\times
(1\pm0.85\delta)\times\sigma_{free}^{BB,inelastic}.\label{inelastic}
\end{equation}
In Eqs.~(\ref{elastic})
and (\ref{inelastic}), whether in initial or final states, ``+'' is for
neutron-neutron scattering while ``-'' for proton-proton scattering.
We neglect the isospin dependent factor in other baryon-baryon scattering cases.

\section{Results and discussions}

\begin{figure}[th]
\centering
\includegraphics[width=0.5\textwidth]{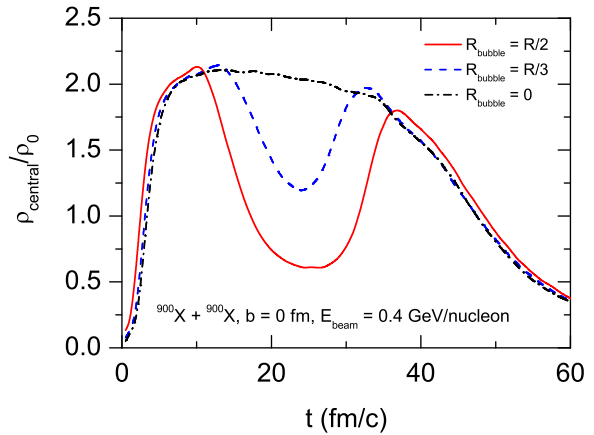}
\caption{(Color online) Evolution of the central density in $^{900}_{274}$X + $^{900}_{274}$X
head-on collisions at a beam energy of 0.4 GeV/nucleon with bubble radii $R_{bubble}$ = 0, $R/2$, $R/3$.}
\label{cden}
\end{figure}
\begin{figure}[th]
\centering
\includegraphics[width=0.5\textwidth]{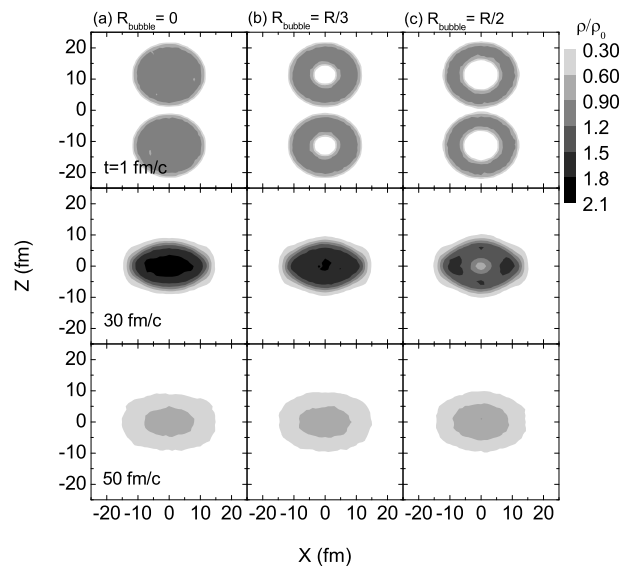}
\caption{(Color online) Evolution of the contour plots of
density distribution in $^{900}$X + $^{900}$X
head-on collisions at the beam energy of 0.4 GeV/nucleon in X - Z
plane with bubble radii $R_{bubble}$ = 0, $R/2$, $R/3$.}
\label{cxz}
\end{figure}
\begin{figure}[th]
\centering
\includegraphics[width=0.5\textwidth]{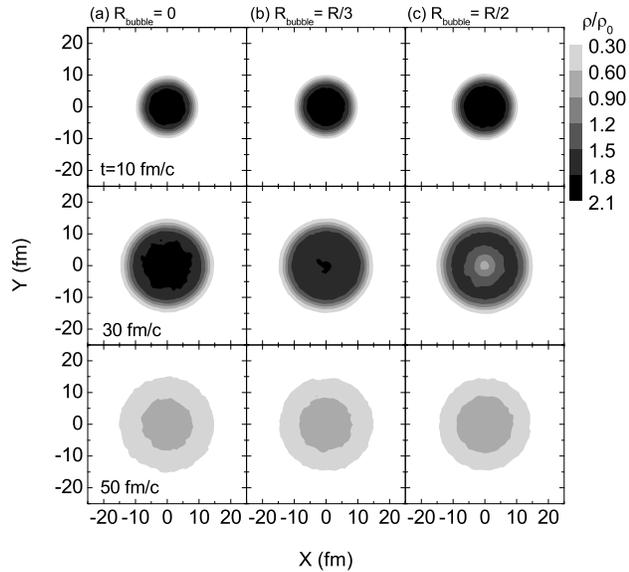}
\caption{(Color online) Same as Fig.~\ref{cxz}, but in X - Y plane.}
\label{cxy}
\end{figure}
%
%\begin{figure}[th]
%\centering
%\includegraphics[width=0.5\textwidth]{fig4.eps}
%\caption{(Color online) Same as Fig.~\ref{cxy}, but with the bubble radius $R_{bubble}$ =
%0.} \label{cxyn}
%\end{figure}
Before studying how to probe the bubble configuration in nucleus in heavy-ion collisions, it is
instructive to show the density-evolution in collisions of
normal and bubble nuclei. Shown in Fig.~\ref{cden} is the evolution of
the central density ($\rho_{central} = \rho (0,0,0)$ in center of mass system with the volume of 1 $fm^{3}$) in $^{900}_{274}$X + $^{900}_{274}$X
head-on collisions at a beam energy of 0.4 GeV/nucleon with bubble radii $R_{bubble}$ = 0, $R/2$, $R/3$. It is clearly seen that, compared with the collision of normal nuclei, there is a serious depletion of central density with bubble configurations in projectile and target nuclei.
Shown in Fig.~\ref{cxz} is the evolution
of contour plots of density distribution in
$^{900}_{274}X$ + $^{900}_{274}X$ head-on collisions at the
beam energy of 0.4 GeV/nucleon in X - Z plane (the beam axis
is in the Z direction). We can see that the maximum compression-%
density region ($\rho/\rho_{0} > 1.8$) is larger for the normal
nuclei while it is smaller if there are bubble configurations
in nuclei. And the larger the bubble is, the smaller the
maximum compression region is seen. This is understandable
since with bubbles in colliding nuclei, the collision of the two
nuclei becomes two bubble's collision, the compression is thus not
strong. It is interesting to see that, if there are larger bubble
configurations in the colliding nuclei, there is also a bubble configuration in the
formed compression matter. Fig.~\ref{cxy} shows the evolution of the
contour plots of density distribution in X - Y plane. Again, it is seen that there is larger compression if there are no bubble configurations
in projectile and target while the compression is smaller if there are bubble configurations
in initial colliding nuclei.

\begin{figure}[th]
\centering
\includegraphics[width=0.5\textwidth]{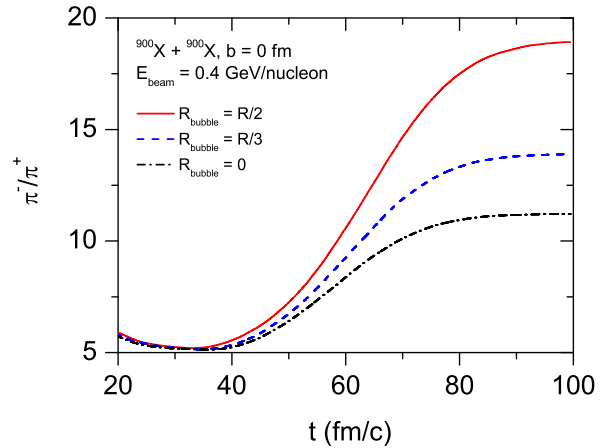}
\caption{(Color online) Evolution of the observable
$\pi^{-}/\pi^{+}$ ratio in the same $^{900}$X +
$^{900}$X head-on collisions ($\gamma$ = 0.3) at the beam energy of 0.4
GeV/nucleon with the bubble radii $R_{bubble}$ = 0, $R/2$, $R/3$.}
\label{trpion}
\end{figure}
\begin{figure}[th]
\centering
\includegraphics[width=0.5\textwidth]{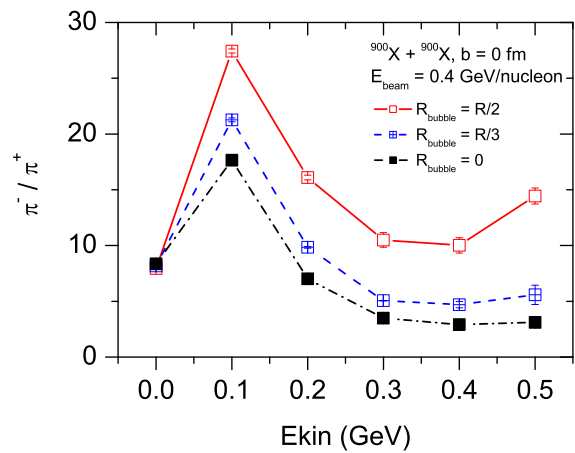}
\caption{(Color online) Same as Fig.~\ref{trpion}, but for the kinetic-energy distribution of the $\pi^{-}/\pi^{+}$ ratio.}
\label{krpion}
\end{figure}
\begin{figure}[th]
\centering
\includegraphics[width=0.5\textwidth]{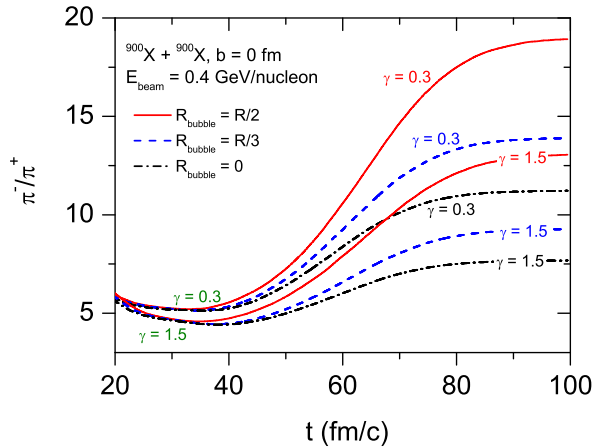}
\caption{(Color online) Effects of the symmetry energy on the
evolution of the observable $\pi^{-}/\pi^{+}$ ratio in the same
$^{900}$X + $^{900}$X head-on collisions at the
beam energy of 0.4 GeV/nucleon with the bubble radii $R_{bubble}$ =
0, $R/2$, $R/3$.} \label{trpion2}
\end{figure}
Since the bubble configurations in colliding nuclei affect maximum compression in heavy-ion collisions and the ratio of $\pi^{-}/\pi^{+}$ is in fact affected by the neutron to proton ratio of compression matter \cite{lyz05}, it is thus naturally to think whether the observable
$\pi^{-}/\pi^{+}$ ratio is affected by the bubble configurations in
colliding nuclei or not. Shown in Fig.~\ref{trpion} is the time
evolution of the observable $\pi^{-}/\pi^{+}$ ratio in the
$^{900}_{274}$X + $^{900}_{274}$X head-on collisions at the
beam energy of 0.4 GeV/nucleon with bubble radii of $R_{bubble}$ =
0, $R/2$, $R/3$. One can clearly see that the value of
$\pi^{-}/\pi^{+}$ ratio from the bubble nuclei's collision is
evidently larger than that from the collisions of normal nuclei.
The effects of bubble configurations in colliding nuclei on the
value of $\pi^{-}/\pi^{+}$ ratio reach about 23\% and 68\%,
respectively, for $R_{bubble}$ = $R/3$, $R/2$. Because the size of the
compression region from the bubble nuclei's collision is smaller
than that from the normal nuclei, the symmetry potential at supra-saturation densities would have less time to act on nucleons. The effect of the symmetry
energy at supra-saturation densities is thus smaller in the collision of the bubble
nuclei's collision. Therefore neutrons are less repelled by the
symmetry potential at high densities in the collision of bubble nuclei. And the
value of the $\pi^{-}/\pi^{+}$ ratio reflects the ratio of neutron number and proton number in
dense matter, i.e., $\pi^{-}/\pi^{+} \approx (N_{dense}/Z_{dense})^{2}$  \cite{lyz05},
it is not surprising to see larger value of the
$\pi^{-}/\pi^{+}$ ratio in the collision of bubble nuclei.

Shown in Fig.~\ref{krpion} is the kinetic energy distribution of
the observable $\pi^{-}/\pi^{+}$ ratio in the same collisions as Fig.~\ref{trpion}.
It can be seen that, for the energetic pion mesons (mainly from
more denser matter), the effects of the bubble configurations on the value
of $\pi^{-}/\pi^{+}$ ratio reach about 70\% and 300\%,
respectively, for $R_{bubble}$ = $R/3$ and $R/2$. Such large
bubble effects on the $\pi^{-}/\pi^{+}$ ratio could be used
to probe the bubble configuration in nucleus.

While in fact the effects of the nuclear symmetry energy on the
$\pi^{-}/\pi^{+}$ ratio in the collisions for heavy nuclei, especially for super-heavy or
hyper-heavy nuclei, are larger than that for lighter nuclei. This is
because the heavier nuclei generally have large $N/Z$ ratio.
Shown in Fig.~\ref{trpion2} is the effects of the
symmetry energy on the evolution of the observable
$\pi^{-}/\pi^{+}$ ratio in the $^{900}_{274}$X +
$^{900}_{274}$X head-on collisions at the beam energy of 0.4
GeV/nucleon with bubble radii of $R_{bubble}$ = 0, $R/2$, $R/3$.
Because the stiffer symmetry energy ($\gamma$ = 1.5) causes more
neutrons to emit from the formed dense matter in collision and the
neutron-poor dense matter causes a small value of the
$\pi^{-}/\pi^{+}$ ratio \cite{lyz05}, as expected, one sees in
Fig.~\ref{trpion2} the value of $\pi^{-}/\pi^{+}$ ratio with the
stiffer symmetry energy ($\gamma$ = 1.5) is smaller than that with
the soft symmetry energy ($\gamma$ = 0.3). Therefore, the effect of
the symmetry energy is one of the largest uncertain factors in
probing the bubble configuration in heavier nuclei by
the observable $\pi^{-}/\pi^{+}$ ratio. Fortunately, the
nuclear symmetry energy at high densities has recently been roughly
pinned down, i.e., a soft symmetry energy is
supported by the FOPI and FOPI-LAND experiments \cite{yong20152}.

\section{Conclusions}

Based on the isospin-dependent nuclear transport model at
intermediate energies, it is found that the bubble
configurations in colliding nuclei affect maximum compression
in heavy-ion collisions. Compared with the collision
of normal nuclei, there is a serious depletion of central
density with bubble configurations in projectile and target
nuclei. Since the symmetry energy plays important role, the observable
$\pi^{-}/\pi^{+}$ ratio in heavy-ion collisions
is sensitive to the bubble configurations in
colliding nuclei. The ratio of $\pi^{-}/\pi^{+}$, especially its kinetic-energy distribution, could be used to probe the bubble configuration in nucleus.

However, the detection of the variation of the value of the $\pi^{-}/\pi^{+}$ ratio with the size of the bubble depends on the experimental ability of discerning bubble nucleus. Furthermore, since the variation of the value of the $\pi^{-}/\pi^{+}$ ratio can be due to other factors (fragmentation of participant nuclei, inelastic cross section of nucleon-nucleon scattering, etc.), from an enhancement of the value of the $\pi^{-}/\pi^{+}$ ratio, one can not deduce the existence of bubble configuration in nucleus before reducing theoretical uncertainties.

\section*{Acknowledgements}

The work was carried out at National Supercomputer Center in
Tianjin, and the calculations were performed on TianHe-1A. The
work is supported by the National Natural Science Foundation of
China under Grant Nos. 11375239, 11435014.

\end{document}